# Mass-extinction: Evolution and the effects of external influences on unfit species

M. E. J. Newman
*Cornell Theory Center, Cornell University, Ithaca, NY 14853–3801.*

B. W. Roberts
*LASSP, Cornell University, Ithaca, NY 14853–2501.*

We present a new model for extinction in which species evolve in bursts or 'avalanches', during which they become on average more susceptible to environmental stresses such as harsh climates and so are more easily rendered extinct. Results of simulations and analytic calculations using our model show a power-law distribution of extinction sizes which is in reasonable agreement with fossil data. We also see a number of features qualitatively similar to those seen in the fossil record. For example, we see frequent smaller extinctions in the wake of a large mass extinction, which arise because there is reduced competition for resources in the aftermath of a large extinction event, so that species which would not normally be able to compete can get a foothold, but only until the next cold winter or bad attack of the flu comes along to wipe them out.

## I. INTRODUCTION

Of all the species that have lived on the Earth since life first appeared here 3 billion years ago, only about one in a thousand is still living today. All the others became extinct, typically within ten million years or so of their first appearance. This high extinction rate has had an important influence on the evolution of life on Earth: the population and repopulation of an ecological niche by species after species allows for the testing of a much wider range of survival strategies than the slower process of *phyletic transformation* by which a species gradually adapts to its surroundings. This in turn has contributed greatly to the diversity of life on the planet.

The importance of extinction to the development of life leads us to some crucial questions, the most fundamental of which is this: is extinction a natural part of the evolutionary process, or is it simply a chance result of occasional catastrophes besetting either single species (such as diseases) or larger groups of species (such as changes in the salinity of the sea, or changes in the climate)? Many talented thinkers have offered arguments on either side of this debate (see, for example, the reviews by Raup (1986) and Maynard Smith (1989)). In this paper we suggest that the truth lies somewhere between the two opposing points of view, and present a model demonstrating how the evolutionary process might interact with environmental stresses to produce a distribution of extinctions similar to that seen in the fossil record.

### A. Bad genes or bad luck?

In his excellent account of extinctions in terrestrial prehistory, Raup (1991) has examined the question of whether extinction arises as a natural part of the evolution process. In his words, do species become extinct through "bad genes or bad luck"? By "bad genes" Raup means extinctions which occur because species are poorly adapted to their surroundings and so have low reproductive success (Hallam 1990). Recently, an interesting new mechanism for "bad genes" extinction has highlighted in the work of Bak and Sneppen (1993), where initially well-adapted species become less well-adapted if one or more of the other species with which they interact (for example by predation or by competition) evolve into some other form. Such a change of situation can force a species to evolve itself, with the ancestral species disappearing (a 'pseudoextinction' in Raup's nomenclature), or it can eliminate a species altogether, leaving available for repopulation by another species the ecological niche that the species used to occupy. In the model proposed by Bak and Sneppen, these changes produce a 'domino' effect in which the evolution of one species causes a number of others to evolve, and they affect others still, and there results an 'avalanche' of evolution propagating through the ecosystem. They suggest that this mechanism alone could be sufficient to explain the mass extinctions seen in the fossil record; extinction could be a natural result of the way in which species evolve, and is not necessarily dependent on external physical factors.

There is another camp however, who point out that there are extinction events in the history of the Earth which have known external causes (Raup and Boyajian 1988), and therefore that the model of Bak and Sneppen is at best incomplete. To take the most famous example, there is now a very convincing accumulation of evidence to suggest that the Cretaceous–Tertiary (K–T) boundary extinction was caused by the impact of a meteor or comet about ten kilometers in diameter on the Yucatán Peninsula in Mexico (Alvarez *et al.* 1980, Swisher *et al.* 1992, Glen 1994). Bak and Sneppen themselves make this point, stressing that the mechanism of their model is not the only one for extinction, but merely that, in the absence of other mechanisms, theirs might still give rise to mass extinction events.

In this paper we propose a new model, similar in many respects to that of Bak and Sneppen, which combines the effects of bad genes and bad luck to make new predictions about the distribution of extinction sizes. The idea behind the model is that species undergo evolution in bursts, as in the model of Bak and Sneppen, and that during these bursts the species will on average be less resilient and well-adapted to their environment than they are at other times. If an external stress is placed on the ecosystem during such a period of evolutionary activity, we therefore expect the extinction rate to be higher than it might have been if the same stress had occurred during a period of relative phenotypic stability. (This idea is not proposed here for the first time—a number of authors have suggested similar mechanisms, e.g., Quinn and Signor (1989), Kauffman (1991, 1993), Plotnick and McKinney (1993).) From this simple hypothesis we have created a model that shows many features seen in the fossil record. We see sporadic bursts of evolution, which Sneppen and co-workers (1994) have likened to the 'punctuated equilibrium' behavior postulated for individual species by Eldredge and Gould (1972, Gould and Eldredge 1977, 1993). We see a (power-law) distribution of extinction sizes ranging from 'mass' extinctions wiping out a significant fraction of all species, to 'background' ones wiping out just one or two species. We see 'precursor' extinctions in which species are seen to be slowly dying off for a certain period before a major extinction event, and 'aftershocks' in which opportunistic, but not particularly well-adapted, species are quickly extinguished as they rise up in the aftermath of a major event.

In Section II we describe our model in detail. In Section III we give the results of our simulations and analytic calculations on the model. In Section IV we give our conclusions.

## II. THE MODEL

Our model is a generalization of the 'minimal self-organized criticality' model for evolution of Bak and Sneppen (1993). (See also Ray and Jan (1994) and de Boer *et al.* (1994).) We consider an ecosystem with a fixed number $N$ of species or species groups, and for the purposes of the model characterize each by just two real numbers: a fitness and a barrier to mutation, denoted $F_i$ and $B_i$ respectively for the $i^{\text{th}}$ species. We use the fitness as a measure of how susceptible a species is to extinction from environmental effects such as climate change, and not as a gauge of the relative merits of one species over another in direct competition between species. There is no mechanism within our model for direct inter-species competition. There is no absolute scale for the fitness measure. For convenience we allow it to take values between zero and one.

The barrier to mutation is a measure of how far a species must mutate against a selection gradient (Caswell 1989) before reaching the domain of attraction of a new evolutionarily stable phenotype. This concept is illustrated in Figure 1, which portrays a section of a 'rugged fitness landscape' (Wright 1982, Kauffman 1993), in which different points on the horizontal axis represent different phenotypes, and the vertical axis measures, for example, lifetime reproductive success, or some other suitable measure of species success. Species spend long periods of time at maxima in this landscape, where they are well-adapted to their environment. Small mutations away from such a maximum are always driven back to the maximum again by the selection gradient. On very rare occasions a species will undergo a large mutation, or possibly a rapid succession of small ones, which will carry it so far from the current maximum that it passes one of the barriers into the domain of attraction of a different maximum. It will then be driven towards that new maximum by the selection gradient in that domain, and probably then remain there for some time, again undergoing small fluctuations about its new form. These phyletic jumps can set off a chain reaction of coevolution in other species, giving a burst or 'avalanche' of evolutionary activity.

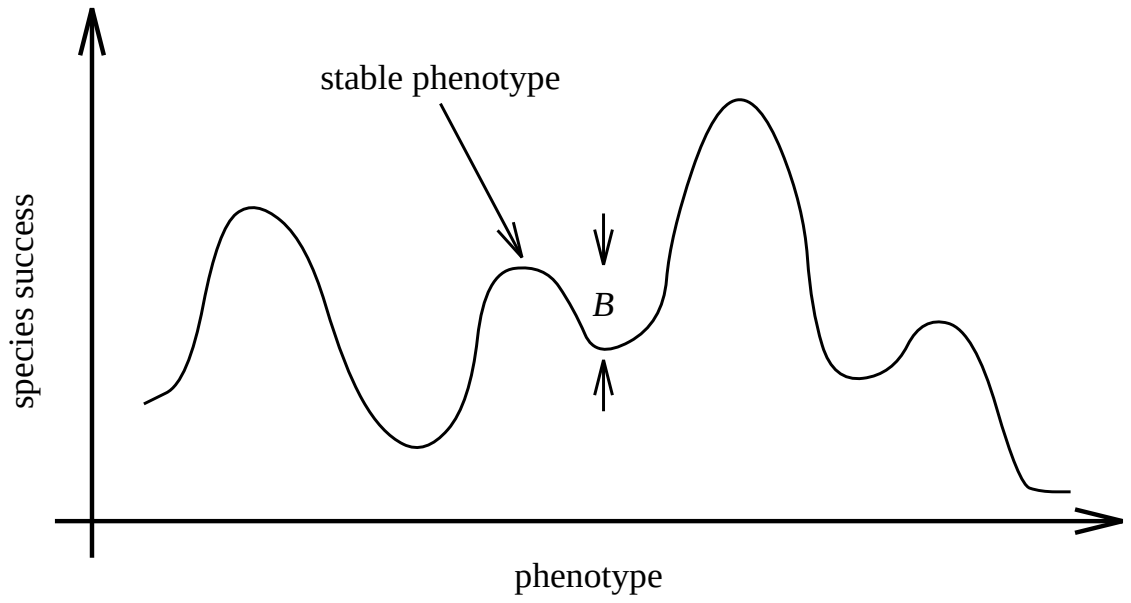

FIG. 1. A schematic representation of a portion of a rugged fitness landscape. Maxima in the fitness function correspond to stable phenotypes. $B$ represents the smallest fitness barrier that has to be traversed by a species at the indicated point in order for it to mutate to a new stable form.

Our barrier variables are a primitive representation of the situation depicted in Figure 1, in which we take into account only the height of the smallest barrier a species needs to traverse in order to reach a new maximum in the fitness landscape. (For the phenotype indicated in the figure, this smallest barrier is shown as $B$.) Again there are no obvious units for the heights of the barriers, so, following Bak and Sneppen, we choose them to lie in the range between zero and one.

Initially we take our $N$ species and assign to each a fitness and a barrier chosen at random within the allowed range. Then we consider how the ecosystem is likely to evolve. Our simulation consists of the repetition, always in the same order, of three basic steps.

The most likely event is that the species with the lowest barrier to mutation—call it species $m$—will evolve first. So the first step is to find that species and have it evolve into some new form characterized by a new value of the fitness $F_m$ and a new barrier to mutation $B_m$, which again we choose at random within the allowed range. This process has the effect of removing the species with the lowest barriers from a population. The second step is to also choose new random values for the fitness and barrier height of $K-1$ 'neighbors' of species $m$. ($K$ is defined in this way for compatibility with the '$NK$' models of Kauffman (1993) and others.) The neighbors are those species with which species $m$ interacts in some fundamental way, for example by predation or by competition for resources. The neighbors can be chosen in a number of different ways. Here we follow Bak and Sneppen and make the simplest choice, whereby the neighbors of species $m$ are selected at random from the $N-1$ possible candidates. The changing of the fitnesses and barriers for the neighbors models the change in the environment of the neighboring species brought on by the change in species $m$ which it undergoes when it evolves. It is this change in the parameters describing the neighbors that gives rise to coevolutionary avalanches within this model, since it may well be that the new barrier to mutation chosen for one of the $K-1$ neighbors will be low enough that it will be the next species that is chosen to evolve, precipitating a chain reaction.

The third and last step in our simulation mimics the effect of environmental forces on our ecosystem. We imagine that environmental forces put some stress on the system which will cause some species to become extinct. Most of the time this stress will not be very severe and the majority of species will be unaffected. But occasionally there will be some larger event which will cause greater extinction. To model such processes we choose a random number $r$ between zero and one at each step in the simulation. This number represents the stress being placed on the ecosystem by external forces. We then assume that all species whose fitnesses are less than this number become extinct at this time, and we replace their fitness and barrier parameters by new ones chosen at random, to represent the repopulation of their ecological niches by newly appearing species. We have experimented with a number of different forms for the random numbers (or 'noise') in our model, including Gaussian (white) noise, $1/f$ noise, exponentially-distributed random numbers, and bimodally-distributed random numbers. The most important predictions of our model are independent of the form of the noise we choose, implying that it is not necessary to know the exact nature or even the cause of the stresses placed on the ecosystem, or their distribution and frequency, in order for the model to make predictions about mass extinction. Most of the results presented here are for Gaussian noise centered at zero with a standard deviation which we denote $\sigma$. In the limit $\sigma = 0$ in which the noise vanishes, the fitness parameters no longer have any effect on the model, since no species ever have a low enough fitness to get wiped out. In this case our model becomes the same as that of Bak and Sneppen. In most of our simulations, we have kept the noise level quite small, with typical values for $\sigma$ being around 0.1 or 0.2.

Our simulations consist of repeating the above processes—evolution of the species with the lowest barrier to mutation, changing of the fitnesses and barriers of its neighbors, extinction of the species with the lowest fitnesses—many times over (typically about $1000N$ times) and examining the resulting pattern of extinctions, the distribution of extinction sizes, and the distributions of fitnesses and barriers to mutation.

## III. RESULTS AND DISCUSSION

We have performed simulations of our model for up to $N = 10000$ species and up to ten million time-steps. (We have not said how long an interval of geological time our time-step corresponds to, but none of the results presented below depend on knowing this. In theory it may be possible to answer this question by comparison of fossil data on species lifetimes with similar data extracted from the simulation. However, we have not attempted to perform this comparison.)

Our simple model where each species interacts with $K-1$ others is only an approximation to the behavior of a real ecosystem; real species can interact with other species strongly or weakly, and different species will interact with different numbers of neighbors. Neither of these effects is allowed for in this simple model. However, if one is to choose a single figure for the number of neighbors for a simulation such as this, then experimental data (for example, the work of Sugihara, Schoenly, and Trombla (1989) on analysis of food webs) suggest a value of 3 or 4. The results presented here are for three neighbors ($K = 4$).

Figures 2 and 3 show the distributions of fitnesses and barriers for two different strengths of the external noise. The individual symbols are the values calculated from our simulations. The lines running through them are the values of the same quantities calculated from an analytic 'mean field' solution of the model. This mean field solution is a generalization of the one given by Flyvbjerg *et al.* (1993) for the model of Bak and Sneppen. The technical details of the solution are to be published elsewhere (Roberts and Newman, in preparation). For the moment we simply note that the mean-field solution agrees excellently with the simulation results. This allows us to probe the behavior of the model in regimes in which the statistics from the simulations are less good, and also in theory to extract exact results for certain quantities appearing the model, such as the exponent $\alpha$ defined in Equation 1.

The distribution of fitnesses shows essentially the behavior we would expect of it: the external stresses placed on the system remove species with low fitness, so that the distribution has most of its weight in the upper half of the range. Towards the top of the range, where noise events large enough to wipe out the species are rare, there is no practical difference between species with different fitnesses, so the distribution is flat. In the limit where the fitness $F \longrightarrow 0$ no species can survive, since there is some small finite noise level present at every step in the simulation, so the distribution is guaranteed to tend to zero in this limit.

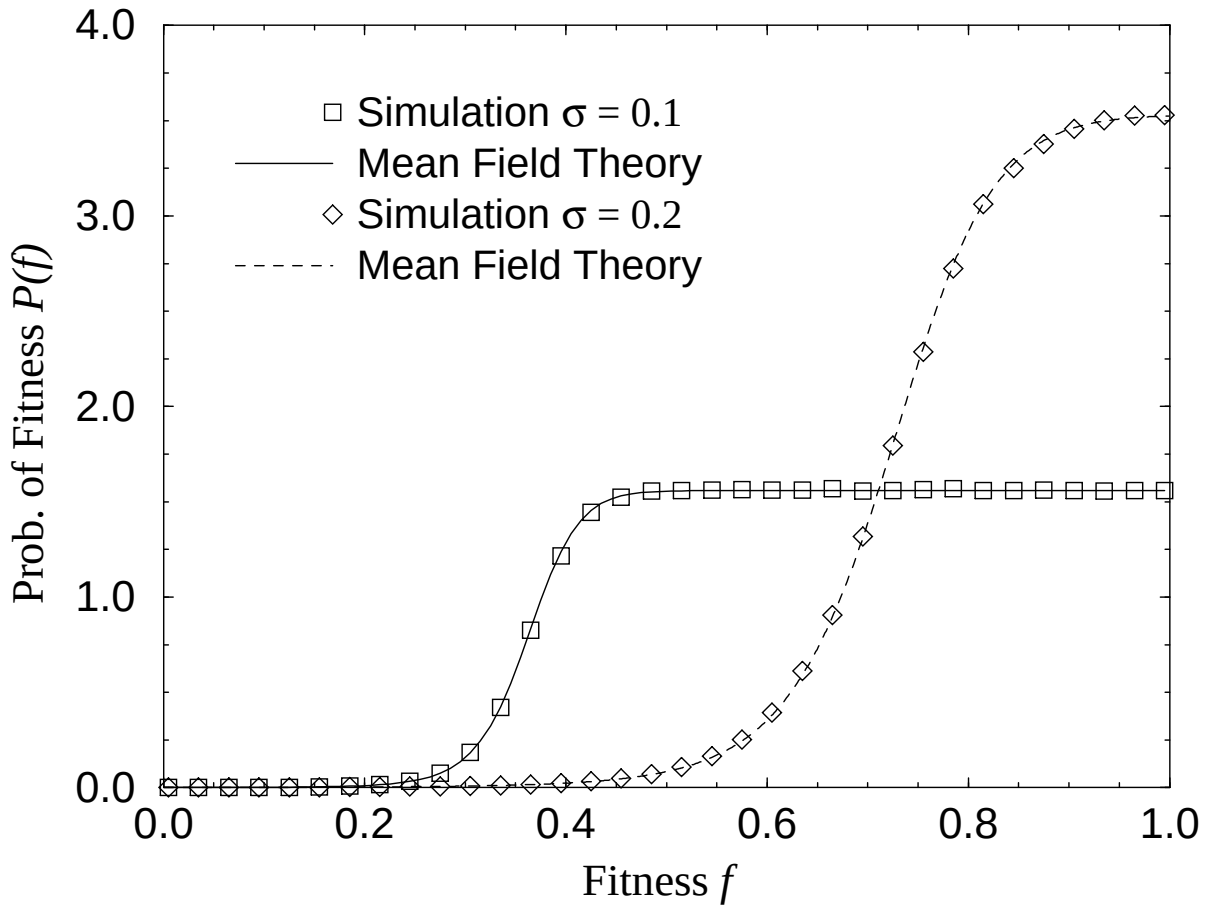

FIG. 2. A histogram of the mean fitness distribution over the whole ecosystem for two different strengths of Gaussian noise. The symbols are the results from the numerical simulations, and the solid lines are the mean-field solution.

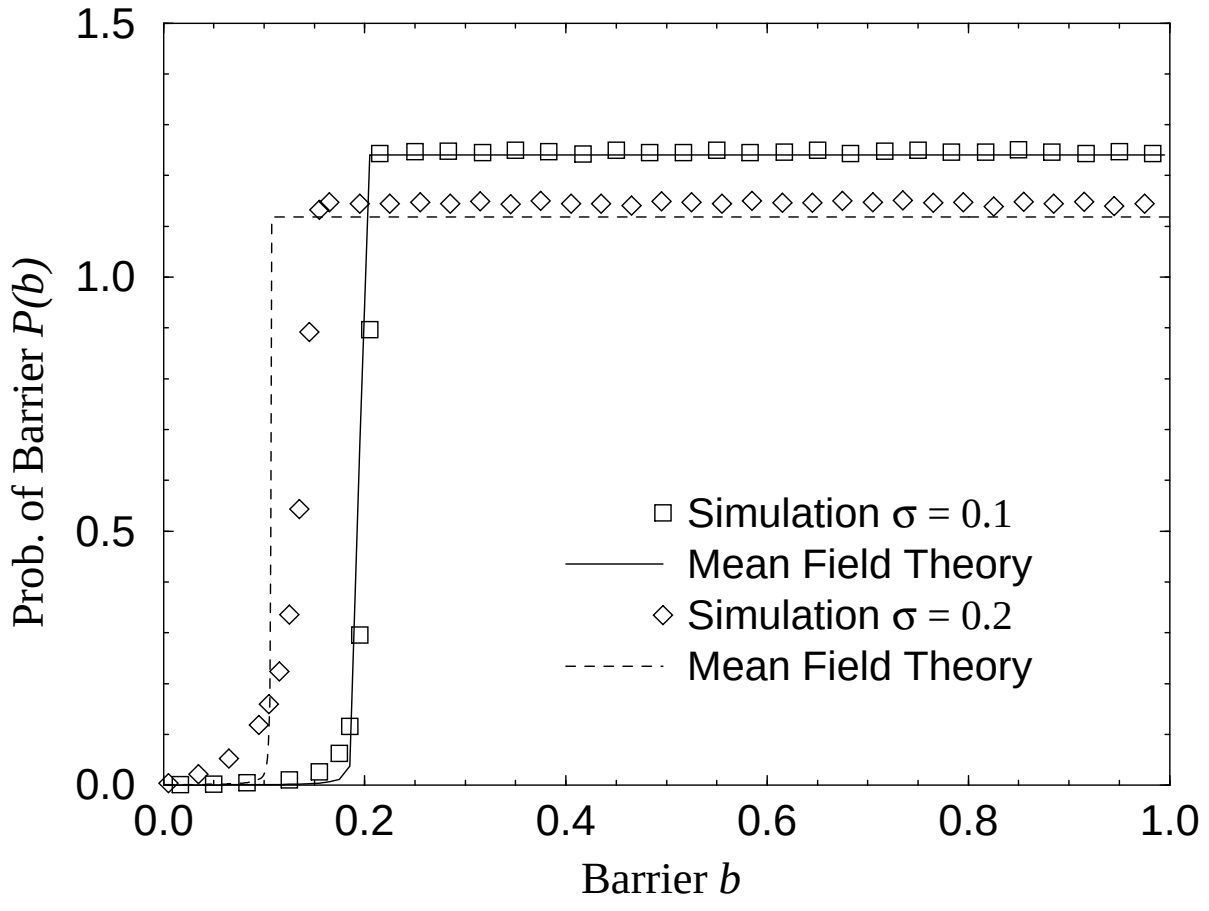

FIG. 3. A histogram of the mean distribution of barriers for two different strengths of Gaussian noise. The symbols are the results from the numerical simulations, and the solid lines are the mean-field solution.

The barrier distribution also tends to zero as barrier height tends to zero, since an infinitesimal barrier is extremely likely to be the lowest one in the ecosystem, causing the corresponding species to evolve to a new form with a different barrier to mutation. In the limit of large barriers, the distribution is again constant, since a large barrier is very unlikely to be the lowest one and there is then no practical distinction between the species at the high end of the distribution.

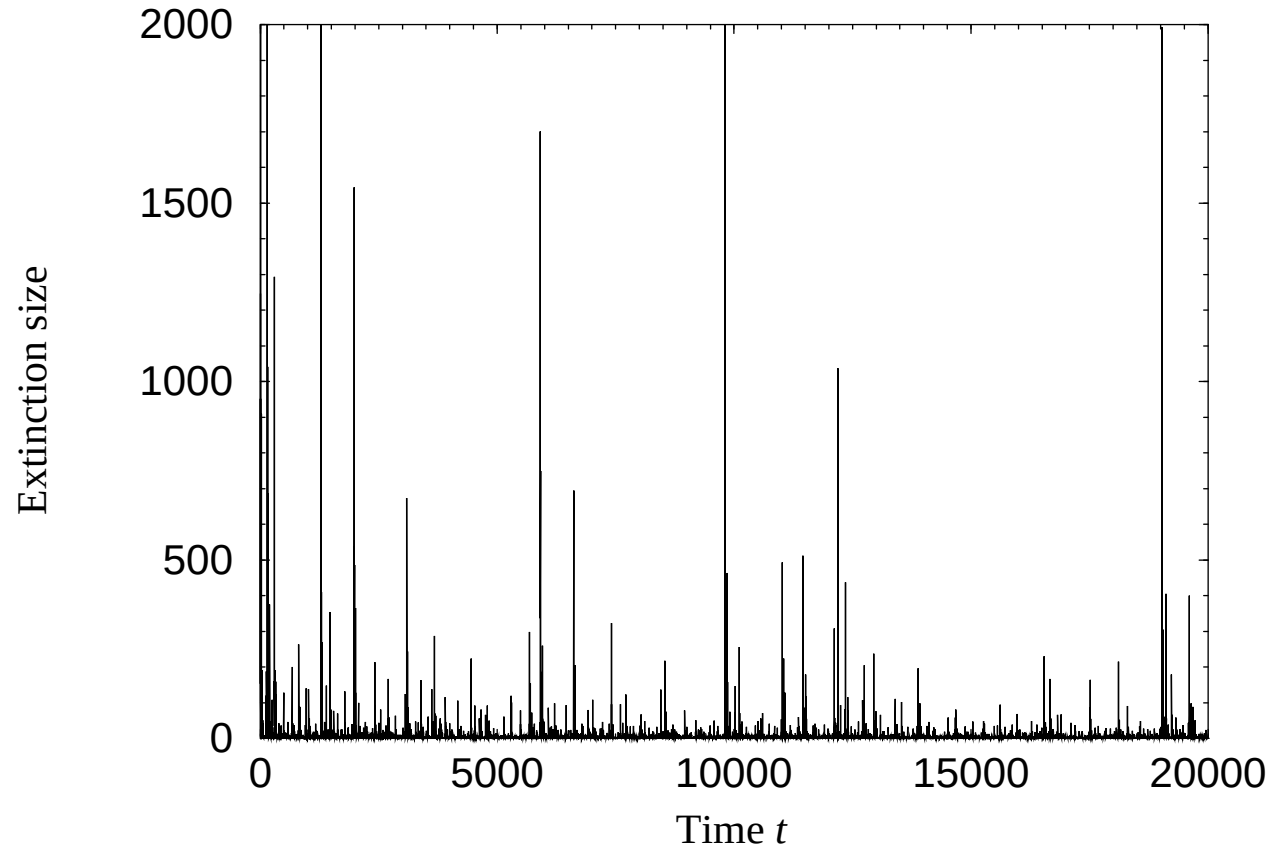

FIG. 4. A section of the extinction data from a simulation of our model with $N = 10000$ and $K = 4$. Notice the punctuated behavior of the model, with long periods of inactivity separated by brief bursts of heavy extinction.

Figure 4 shows a plot of the extinctions occurring in a population of 10000 species against time. There are long periods during which there is little activity—a few species dying out here and there, but nothing particularly catastrophic—followed by a few large extinction events, followed by a further period of relative inactivity, and so forth. To understand the mechanism through which this behavior arises within the model, consider the behavior of species' fitnesses as time goes by. Starting with a well-adapted ecosystem in which the bulk of the species have a high fitness (tolerance to external influences), we allow our process of evolving the species with the lowest barrier to proceed for a while. At each time-step this replaces $K$ species with new ones with randomly chosen barriers and fitnesses. There is a reasonable chance that these fitnesses will not be as high as those of the well-adapted species from which these new ones have evolved, and so these species may be wiped out by quite small environmental stresses. These are the small extinctions we see going on most of the time in Figure 4.

In fact, the evolutionary process will proceed, as described in Section II, in coevolutionary avalanches, most of which will be small. However, as shown by Bak and Sneppen (1993) there is a power-law distribution of these avalanches, and occasionally large ones occur. When this happens, a large number of species have their fitnesses changed to new random values, and consequently a large number become more susceptible to external factors. As long as the noise level remains low, this is not a problem, but if a particularly large noise event occurs, then a significant fraction of these species can be wiped out, giving rise to the large extinction events seen in the data. The important point here is that large extinctions arise as a result of the coincidence of catastrophic environmental changes with large coevolutionary avalanches, during which the susceptibility of large numbers of species to external effects is increased.

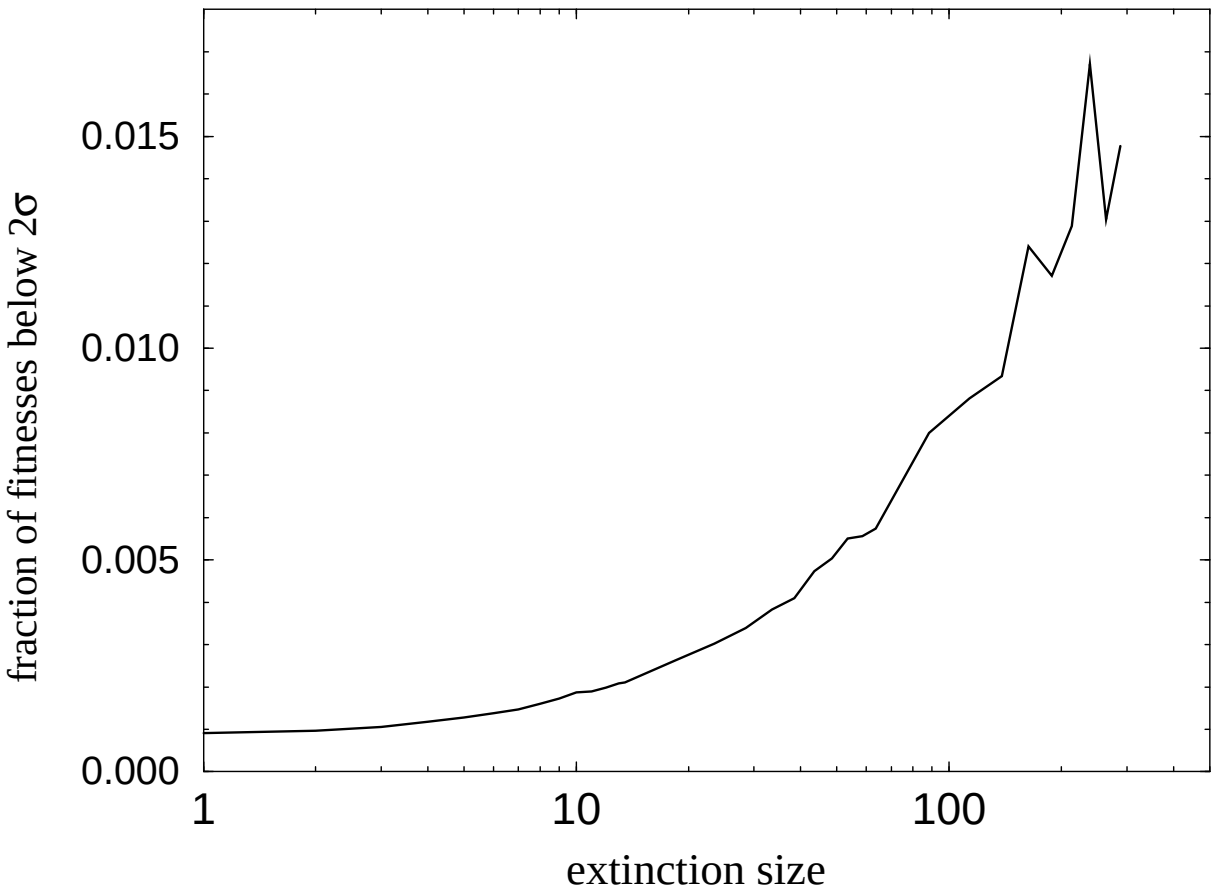

FIG. 5. The fraction of the species having fitnesses below a certain threshold (twice the standard deviation of the noise in this case) immediately before extinctions of a certain size. Notice that the number of species having low fitness is higher immediately before a larger extinction, indicating that the large extinctions are the result of the coincidence of lower fitness with large environmental stresses.

There are many indicators in our simulation results that this is the correct explanation of the observed distribution of extinctions. For instance, we have calculated the fraction of the species which have fitness below a certain threshold immediately before each avalanche (in the present case we took that threshold to be twice the noise level), and then taken the average of this quantity for each size of avalanche. The results of this calculation are shown in Figure 5, and it is clear that the fraction of species with low fitnesses increases before a large avalanche. Large extinctions are therefore not just an effect of large environmental changes.

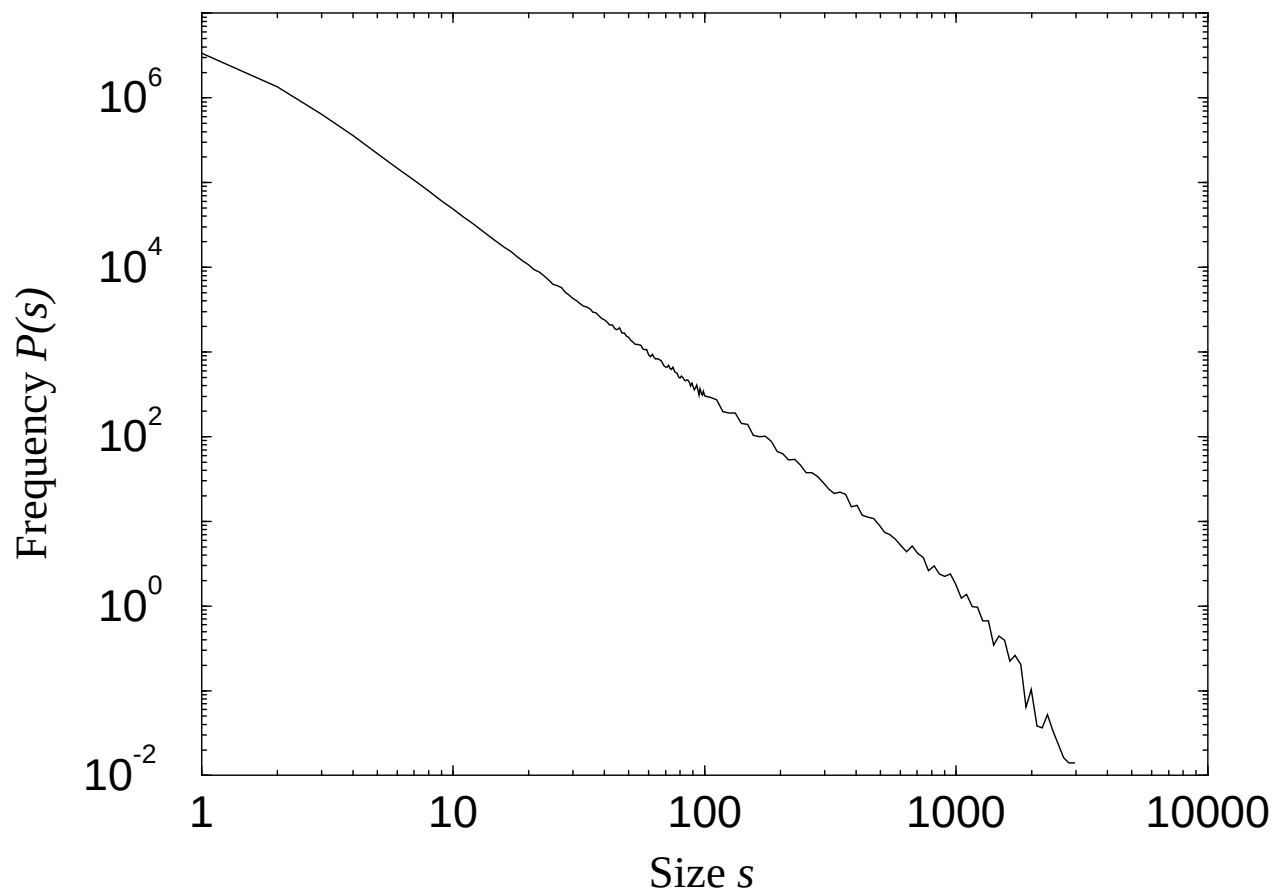

FIG. 6. Log-log plot of the distribution of extinction sizes in the model. The straight-line form of the graph indicates that the distribution is a power-law, and the gradient of the line gives a value $2.183 \pm 0.007$ for the exponent of the power-law.

Another important indication is shown in Figure 6, which is a logarithmic histogram of the size of extinctions against the frequency of occurrence. The first thing we notice is that the graph goes steeply downwards, indicating that large extinctions are much less common than small ones. We also notice that on this plot the distribution falls on a straight line for most of its range, indicating that the extinction distribution is a power law. In fact we find that the frequency of an extinction $P(s)$ is related to its size $s$ by

$$P(s) \propto s^{-\alpha}, \tag{1}$$

where

$$\alpha = 2.183 \pm 0.007.$$

Bak and Sneppen find a power-law distribution for the coevolutionary avalanches in their model, but with an exponent different from the one found here. They equate their avalanches with extinctions, but we suggest that the avalanches, whilst being a crucial part of the mechanism giving rise to mass extinction, do not themselves represent actual extinctions. In fact, we should expect our exponent $\alpha$ describing the distribution of extinctions to be greater than the exponent found by Bak and Sneppen for the avalanches. The reason is that we require the coincidence of two unlikely events—environmental catastrophe and a large coevolutionary avalanche—to produce a large extinction event. Merely having a large avalanche is insufficient; large avalanches take place without giving rise to large extinctions because the environmental conditions are not right for extinction. Thus we expect large extinctions to be relatively less common than large avalanches, and so we expect our power-law distribution to be steeper than that for the avalanches. This is in fact what we observe, since Bak and Sneppen find a value of 1.35 for the exponent governing their power-law distribution, which is considerably less than our 2.18.

The form of the power-law distribution is one of the most robust predictions of our model. We have run the model with a variety of different types of noise, with different numbers of neighbors $K$, and with a number of variations in the precise dynamics of the model, all without changing the value of the exponent $\alpha$ in (1). Thus it is not necessary to know exactly what external effects are responsible for the mass extinctions, or what their distribution is over time—it makes no difference to this particular prediction. This is an important observation since it should, in theory, be possible to check this prediction against paleontological data, for example the fossil data of Sepkoski (1993) or of Williamson (1981).

Another telling form of behavior seen in our model is the generation of smaller extinctions that accompany the largest ones. These fall into two classes.

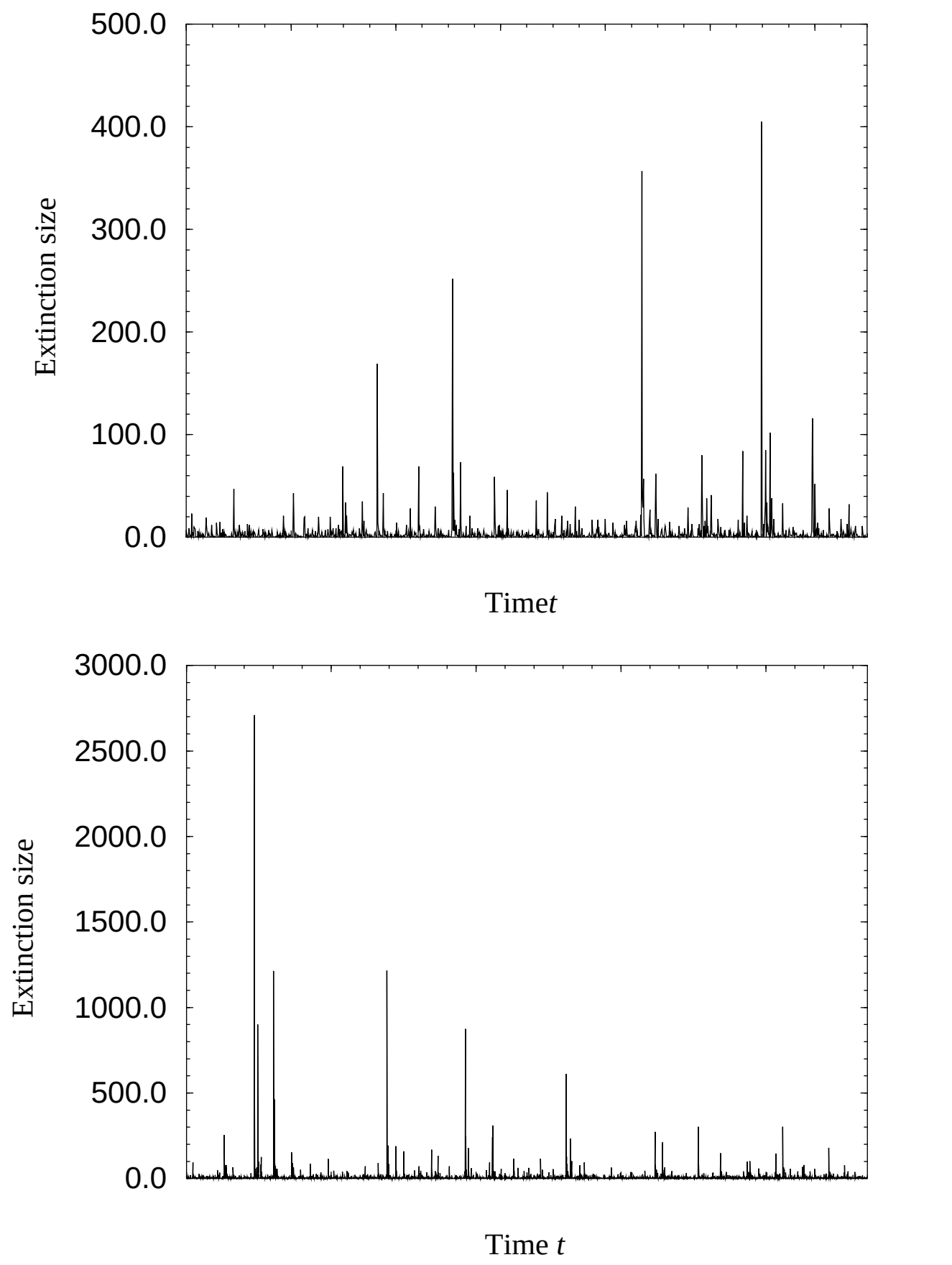

FIG. 7. (a) An example of the 'precursor' effect described in the Section III. (b) An example of the 'aftershocks' described in Section III.

*Precursors* is the name we give to sets of small extinctions which precede a large extinction. Such precursor extinctions are to be seen all over the data from our simulations. Figure 7 (a) shows a set of precursor extinctions drawn from the data shown in Figure 4. The explanation of this effect is as follows. Large extinctions are produced when a large portion of the population becomes less fit than the average following one or more large coevolutionary avalanches. After such an avalanche, the next large environmental stress placed on the system (if there is one before the species involved manage to evolve to a fitter state) will decimate the population. However, there may be a considerable interval of time before such a large stress occurs, and in the meantime, there may be smaller stresses which, because of the general unfitness of the population, tend to exact more of a toll on the ecosystem than one would normally observe. It is this exaggeration of the effect of small environmental changes in the interval following a large avalanche but preceding the next large noise event that we see in our data as precursor extinctions. It is interesting to note that precursor extinctions are seen in the fossil record also. The mass-extinction which occurred at the K–T boundary 65 million years ago was preceded by about three million years during which many species on the planet were already dying out (Keller 1989). The K–T boundary extinction is thought to have been caused by the impact of a large comet or meteor at Chicxelub on the Yucatán Peninsula, and this event corresponds to the 'large noise event' in our theory. However, a comet cannot explain why species were dying out for three million years before, and it has been suggested (for instance, by Kauffman 1993) that these precursor extinctions might be a result of smaller environmental stresses on a population which had become unfit for some other reason. This same unfitness would also go some way to explaining why the population of the planet was so severely reduced by the Chicxelub meteor. It is interesting therefore to observe the exact same effect appearing in our model, with the unfitness here caused by large coevolutionary avalanches.

We give the name *aftershocks* to larger-than-normal extinctions that arise in the wake of a significant mass extinction. The mechanism here is that a large extinction wipes out a significant number of species, leaving empty many ecological niches. These niches are soon filled by new species. However these species may not be very well adapted to survive new environmental stresses, since they have not evolved for long enough to feel the selection pressure of those stresses. Thus many of them will quickly be wiped out by quite small noise events, which normally would have little effect on a well-adapted population. Only with the passage of time can the less fit species be removed from the population and the general fitness increase to normal levels again. Thus we expect to see a series of moderate-sized extinction events appearing in the immediate wake of a large event, dying away in size until we return to the normal spectrum of small extinctions. Figure 7 (b) shows just such a set of aftershocks, also drawn from the data shown in Figure 4.

Again, an effect similar to the aftershocks seen in these simulations appears to be present in the fossil record. For example, during the well-known precambrian explosion of 600 million years ago, a large number of species, many with rather bizarre and ill-adapted morphologies arose very quickly in all sorts of different ecological niches. Most of these were wiped out rather quickly, many probably because they were not well-able to cope with small changes in their environment. Thus extinction, as well as speciation, appears to have been at a maximum during this period. It is interesting to see this effect duplicated in our simulations.

As a last comment on our results, we would like to point out that, although we have experimented with a large variety of different types of noise, to mimic different distributions of external influences, and found essentially the same predictions for the parameters of the system for all of them, there is one important respect in which all of these noise distributions were the same: they were all essentially similar at all points in time. It has been suggested (Raup and Sepkoski 1984, Sepkoski 1990) that there could be some periodicity in the largest of the mass extinctions seen in the Earth's fossil record, caused perhaps by the periodic recurrence of some astronomic catastrophe such as a meteor impact. We have performed simulations of our model which mimic this effect by introducing periodic variation in the strength of the noise function. The resulting extinctions show clear peaks at regular intervals, some more pronounced than others, in a fashion qualitatively similar to the peaks seen at 26 million year intervals in the fossil data. This does not of course prove that there is an external cause for any periodicity that may be present in the fossil extinction data, it merely demonstrates that, within our model at least, such external causes can produce periodicity.

## IV. CONCLUSIONS

We consider a mechanism whereby extinctions might arise as the result of the coincidence of coevolutionary avalanches giving rise to low general fitness of species in an ecosystem, and environmental stresses which tend to wipe out the least fit species. We have given a simple mathematical model which mimics this behavior and makes a number of predictions about the resulting distribution of extinctions. In particular, the extinctions appear to have a power-law distribution with an exponent independent of the precise form of the external stresses. This prediction should be testable against paleontological data.

## V. ACKNOWLEDGEMENTS

We would like to thank H. K. Reeve, K. Sneppen, R. D. Yanai, D. M. Raup, R. E. Plotnick, and R. I. Muetzelfeldt for useful discussions and comments. This work was supported in part by the NSF under grant number DMR–91–18065, and by the Hertz Foundation.